# Proceedings of Workshop AEW10

10th Asia-Europe Workshop on

Concepts in Information Theory and Communications

*A Tribute to Hirosuke Yamamoto*

June 21-23, 2017, Boppard, Germany

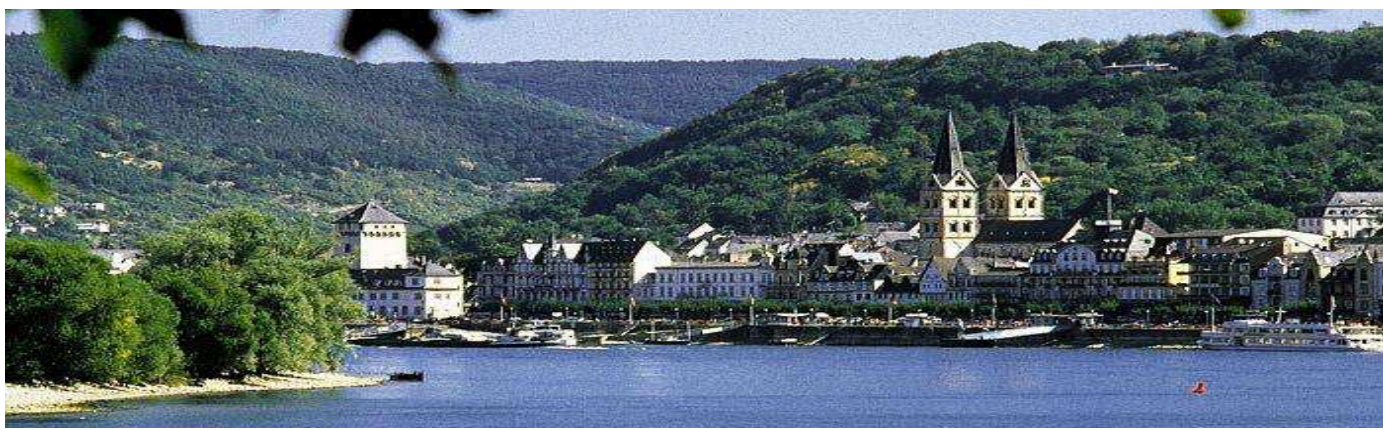

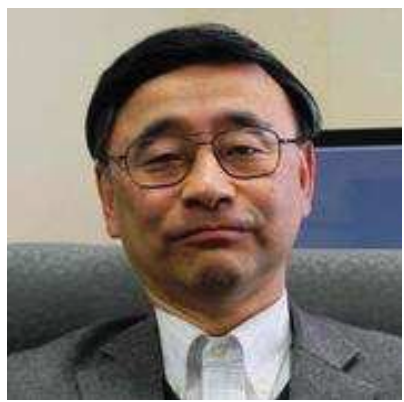 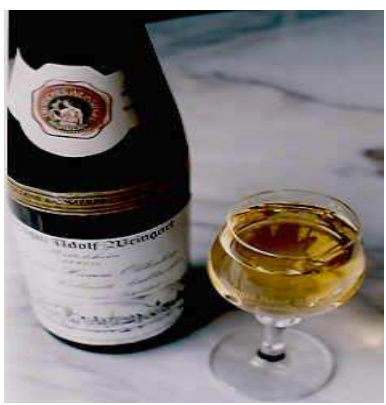 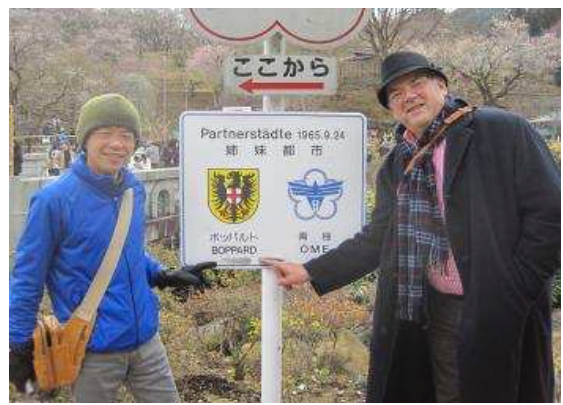

## Sponsored by

**TURING MACHINES INC.**

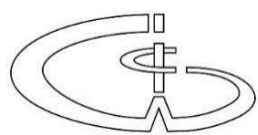

Benelux Working Community on Information Theory.

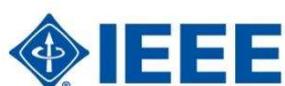 Germany Chapter on Information Theory. 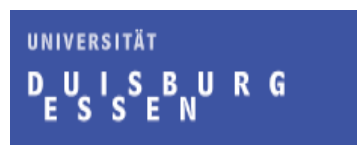

# Proceedings of Workshop AEW10

10<sup>th</sup> Asia-Europe Workshop on

## Concepts in Information Theory and Communications

*A Tribute to Hirosuke Yamamoto*

June 21-23, 2017, Boppard, Germany

## Sponsored by

TURING MACHINES INC.

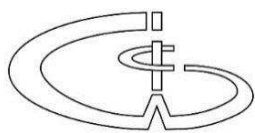
Benelux Working Community on Information Theory.

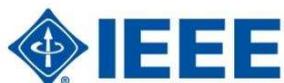
Germany Chapter on Information Theory.

UNIVERSITÄT DUISBURG ESSEN

# Preface

Welcome to Boppard!

The Workshop AEW10 was held in Boppard - Germany on June 21-23, 2017. The theme is "Concepts in Information Theory and Communications".

The 10th Asia-Europe workshop in "Concepts in Information Theory and Communications" is based on a longstanding cooperation between Asian and European scientists. The first workshop was held in Eindhoven, the Netherlands in 1989. The idea of the workshop is threefold: 1) to improve the communication between the scientist in the different parts of the world; 2) to exchange knowledge and ideas; and 3) to pay a tribute to a well respected and special scientist.

We are very proud that for this workshop Hirosuke Yamamoto accepted our invitation to be the guest of honour. We also consider Hirosuke Yamamoto as our key lecturer. His style of presenting is extremely inspiring and always very clear. The organizers want to pay special attention to these aspects. Presenters are requested to pay attention to the educational aspects of their presentation and concentrate on the main ideas and concepts in the developed theory.

A special session "Hardware-Aware Information Theory" will be organized by Brian Kurkoski and Dirk Wbben. A special session on Powerline Communications will be organized by Adel Ghazel and Han Vinck.

The mayor of Boppard, Dr. Walter Bersch, opened the AEW10 and welcomed our guests from Japan. Boppard has a long tradition in the cooperation with Japan, especially the city of Ome.

<div style="text-align:right">A. J. Han Vinck & Hiroyoshi Morita</div>

# Organization

### General co-chairs

Jos Weber ................. Delft University of Technology, the Netherlands
Tadashi Wadayama ........ Nagoya Institute of Technology, Japan

### Program co-chairs

Brian M. Kurkoski ......... Japan Advanced Institute of Science and Technology, Japan
Yanling Chen ............. University of Duisburg-Essen, Germany

### International Advisors

A. J. Han Vinck ............ University of Duisburg-Essen, Germany
Hiroyoshi Morita .......... The University of Electro-Communications, Japan

### Local Arrangements

A. J. Han Vinck ............ University of Duisburg-Essen, Germany

### Proceedings

Yanling Chen ............. University of Duisburg-Essen, Germany
A. J. Han Vinck ............ University of Duisburg-Essen, Germany

# Table of Contents

Note: Abstracts with the roman page numbers are not available.

**Invited Talk**

**Session 5**

**Session 6**

# Progress in Constrained Codes

Kees A. Schouhamer Immink
Turing Machines Inc., Rotterdam, The Netherlands

ABSTRACT

A constrained code is a key component in the digital recording devices that have become ubiquitous in computer data storage and electronic entertainment applications. Early example are runlength-limited codes as first applied in hard disk drives marketed by IBM in the 1950s.

The author surveys the theory and practice of constrained coding, tracing the evolution of the subject from its origins in Shannon's classic 1948 paper to present-day applications in high-density mass data storage systems. Open problems and future research directions are addressed.

# Immink Codes and State Splitting Method

Hiroshi Kamabe
Gifu University, Japan

ABSTRACT

In information theory and communication theory, we usually assume that communication channels can accept any sequences of channel symbols. However, in several applications, it is assumed that the channels accept only sequences satisfying some constraints because input sequences violating the constraints cause malfunctions of the channels. The constraints are called input constraints of the channels.

One of typical examples of input constraints is a $(d,k)$-constraint which requires that the lengths of runs of 0s should be at least $d$ and at most $k$ where 'run' of 0 means a sub-sequence consisting of consecutive 0s and the sub-sequence is taken so that its length is as long as possible. If $d=0$ and $k=\infty$, then the constraint is called a $k$-constraint and $d$-constraint, respectively.

The input constraints which can be represented by finite directed graphs with labeled edges were well studied [1]. Many code construction methods for input constraints were proposed. The state splitting method, proposed by Adler, Coppersmith and Hassner, is one of them and universal in the following sense [2]. If we are given an input constraint represented by a finite directed graph with an injective edge labeling and an $R_0 < C$, then we can construct a finite state encoder with a sliding block decoder and code rate at least $R_0$, where $C$ is the capacity of the constraint. If the edge labeling is not injective, then first we approximate the constraint with a graph having an injective labeling from inside and apply the ACH coding to the approximation of the constraint [3].

Since codes for channels with input constraint are actually used in practical devices, the compactness of the resulting code and performance indices related to specific applications should be taken into account in designing the codes. Hence, coding techniques specific to some input constraints were also proposed. These techniques may produce codes which are more compact than codes produced by the general techniques. K. Immink proposed many techniques for $(d,k)$-constraints and a spectral null constraint. These techniques can be found in [4].

Here we consider a code construction techniques for the $d$-constraint given in [4, Section 7.7]. The techniques usually gives small codes. The technique was applied to phase change memory (PCM) by K. Cai in order to construct codes for a $k$ constraint with very high efficiency [5]. The $k$-constraint for PCM is imposed so that some performance index of the resulting code is optimized. The performance index depends on constrained sequences actually produced by the resulting code. If we use the Immink coding technique in designing the code, we can easily adjust the code by appropriately choosing code words so that the performance index is optimized.

We interpret the Immink coding technique in terms of the state splitting method and characterize the coding technique using follower sets. Then we generalize the coding technique. We also give a constraint to which the coding technique can be applied.

# Periodic-Finite-Type Systems for Flash Memories

Akiko Manada
Graduate School of Informatics and Engineering
The University of Electro-Communications
1-5-1, Chofugaoka, Chofu, Tokyo, 182-8585, JAPAN
Email: amanada@uec.ac.jp

## I. INTRODUCTION

A flash memory is a high-density non-violate memory which has been used in data storage devices such as USB memories or SSDs. Each flash memory consists of cells (floating gates) which accumulate electric charges for storing messages. More precisely, a flash memory stores messages by adding electric charges to the cells, and reads messages by quantizing electric charges stored in the cells. It is easy to increase the charge levels of cells since it can be done by injecting charges per cell. However, it is costly to decrease the charge levels since it must be done per block (a huge set of cells). Hence, many effective coding schemes which considers the constraint on overwriting have been studied (*e.g.*, [1], [2]).

It is also important to handle possible errors of flash memories for accurate message readout. For example, an inter-cell interference (ICI), a shift of charge levels due to the interference of charges in neighbouring cells, is a typical noise for flash memories. Since the likelihood of ICI depends on a data sequence stored in cells, messages should be encoded so that resulting data sequences do not contain subsequences that can cause ICI (constraint on sequences) (*e.g.*, [3], [4]).

Therefore, for producing reliable flash memories, it is desired to consider both constraints into consideration. It implies the necessity of analysis on two-dimensional constrained systems, but the analysis is cumbersome, and many problems have been open up to this moment. We will therefore focus on Periodic-Finite-Type shifts (PFTs), a class of one-dimensional sofic shifts, and discuss how they can be used as a mathematical model of flash memories considering constraints.

## II. DISCUSSION

Let $\Sigma$ be an alphabet, and denote by $\Sigma^*$ and $\Sigma^{\mathbb{Z}}$ the sets of finite sequences (*words*) and bi-infinite sequences over $\Sigma$, respectively. Define the *shift map* $\sigma : \Sigma^{\mathbb{Z}} \to \Sigma^{\mathbb{Z}}$ so that for a bi-infinite sequence $\mathbf{x} = \ldots x_{-1}x_0x_1\ldots$ over $\Sigma$, the $i$-th symbol of $\sigma(\mathbf{x})$ is equal to $x_{i+1}$. For simplicity, let $\sigma^r(\mathbf{x})$ be the bi-infinite sequence generated by applying $\sigma$ $r$ times to $\mathbf{x}$.

A *periodic finite shift* $\mathcal{S}_{\{\mathcal{F},P\}}$ [5], [6] is a set of bi-infinite sequences characterized by period $P$ and an ordered list $\mathcal{F} = (\mathcal{F}_0, \mathcal{F}_1, \ldots, \mathcal{F}_{P-1})$ of finite forbidden sets $\mathcal{F}_j \subset \Sigma^*$, which forbids the appearance of words in a periodic manner. More precisely, a bi-infinite sequence $\mathbf{x} = \ldots x_{-1}x_0x_1\ldots$ is in $\mathcal{S}_{\{\mathcal{F},P\}}$ if and only if there exists an integer $r$, $0 \leq r \leq P-1$, such that $\sigma^r(\mathbf{x})$ does not contain any word in $\mathcal{F}_j$ as a subword starting at the positions $i$ satisfying $i \equiv j \pmod{P}$. PFTs are also sofic shifts; that is, each PFT has its presentation (a labelled directed graph whose bi-infinite labelled paths represent bi-infinite sequences in the PFT).

Consider a block consisting of $n$ $q$-level cells, and suppose that data sequences are overwritten in the block. Let $\Sigma = \{0, 1, \ldots, q-1\}$ and $\ell^{(t)} = (\ell_1^{(t)}, \ell_2^{(t)}, \ldots, \ell_n^{(t)}) \in \Sigma^n$ be the stored data sequence at the $t$-th write; that is, the $i$-th element of $\ell^{(t)}$ represents the charge level of the $i$-th cell at the $t$-th write. Set $\ell^{(0)} = (0, 0, \ldots, 0)$ as initialization.

Fix a prime number $P$. Define $\mathcal{F}^{(t)} = (\mathcal{F}_0^{(t)}, \mathcal{F}_1^{(t)}, \ldots, \mathcal{F}_{P-1}^{(t)})$ to be an ordered list of finite forbidden sets at the $t$-th write, where $\mathcal{F}_0^{(0)} = \mathcal{F}_1^{(0)} = \cdots = \mathcal{F}_{P-1}^{(0)}$ for $\mathcal{F}^{(0)}$. Given $\ell^{(t)} = (\ell_1^{(t)}, \ell_2^{(t)}, \ldots, \ell_n^{(t)})$, compute $a_j^{(t)} = \max\{\ell_i^{(t)} : i \equiv j \pmod{P}\}$ for each $0 \leq j \leq P-1$. Then, update $\mathcal{F}^{(t+1)} = (\mathcal{F}_0^{(t+1)}, \mathcal{F}_1^{(t+1)}, \ldots, \mathcal{F}_{P-1}^{(t+1)})$ based on $\mathcal{F}^{(t)} = (\mathcal{F}_0^{(t)}, \mathcal{F}_1^{(t)}, \ldots, \mathcal{F}_{P-1}^{(t)})$, so that

$$\mathcal{F}_j^{(t+1)} = \mathcal{F}_j^{(t)} \cup \{0, 1, \ldots, a_j^{(t)} - 1\}.$$

If PFT $\mathcal{S}_{\{\mathcal{F}^{(t+1)}, P\}} \neq \emptyset$, we can find a data sequence $\ell^{(t+1)}$ satisfying constraints for the $(t+1)$-th write.

The coding rate at the $t$-th write is lower bounded by the capacity of $\mathcal{S}_{\{\mathcal{F}^{(t)}, P\}}$, which is obtainable from its presentation. However, it requires high computational complexity when $P$ is large, or the restriction of data sequences due to $\mathcal{F}_j^{(t)}$ gets tight when $P$ is very small compared to $n$. Hence, a suitable value on $P$ and relaxed conditions on $\mathcal{F}_j^{(t)}$ should be proposed.

## ACKNOWLEDGMENT

This work was supported in part by JSPS KAKENHI Grant-in-Aid for Young Scientists (B), Grant Number 15K15936.

# Concept of CoCoNuTS

Jun Muramatsu* and Shigeki Miyake**
*NTT Communication Science Laboratories, NTT Corporation (muramatsu.jun@lab.ntt.co.jp)
**NTT Network Innovation Laboratories, NTT Corporation (miyake.shigeki@lab.ntt.co.jp)

## ABSTRACT

In this talk, we introduce the concept of CoCoNuTS (Codes based on Constrained Numbers Theoretically-achieving the Shannon limits). This concept provides building blocks for codes achieving the fundamental limits.

Our study on the construction of codes achieving the Shannon limits has been continued for over ten years. Tools for the code construction has been developed one by one and published in the subsequent papers. This talk reviews and summarizes our results for the construction of a channel code achieving the fundamental limits. The construction of a lossy code is presented in [4][5].

The idea of the construction is originated from the application of the Slepian-Wolf code to channel coding. This idea was first presented in *this workshop* [3, Section III], and published in [14, Theorem 4]. Later, we noticed that the similar idea had already been given by Shannon [15, Fig. 8]. He assumed that an observer has access to both input and output of a channel and send the difference of them to the decoder. In contrast, our idea suggests that such an observer is unnecessary by sharing the common information between the encoder and the decoder. However, there was no practical construction of both an encoder and a decoder.

The first progress is using a two functions (sparse matrices) to construct a code. The effectiveness of this idea was shown in [1][2]. Later, the essence of this idea, which is called *hash property*, was introduced in [6][7], where it is shown that the idea can be applied to the constructions of multi-terminal codes [8][9][11]. However, encoding and decoding were still intractable.

The second progress is using a stochastic encoder called a *constrained-random-number generator* introduced in [4][5], where the theoretical study comes from [10]. Two approximation algorithms using the sum-product algorithm and the Markov Chain Monte Carlo method were developed. This time, we obtained tractable algorithms for encoding but the proof of achievablity to the fundamental limit depends on the use of the maximum-likelihood decoder, which is still intractable.

The third progress is separating encoding and decoding functions. In our above results, we had to find a pair of good functions (sparse matrices) to achieve the fundamental limit. It is shown in [12] that for a given arbitrary function we can construct a code achieving the fundamental limit.

The last progress is using a constrained-random-number generator for the stochastic decoding. The effectiveness of this idea will be presented in [13]. The result implies that both an encoder and a decoder can be constructed by using constrained-random-number generators, where we can use tractable encoding/decoding algorithms by using the sum-product algorithm and the Markov Chain Monte Carlo method.

# Double Nearest-Neighbor Error Correcting Codes

Hiroyoshi Morita
The University of Electro-Communications
Chofu, Tokyo 182-8585, JAPAN
Email: morita@uec.ac.jp

## Abstract

We propose a new class of double nearest-neighbor error-correcting codes which are suited to multi- dimensional signal constellations and present the corresponding decoding algorithm using elementary arithmetics. The proposed code $\mathcal{C}$ is a linear $[n, n-2]$ code over $GF(p)$ where $p \equiv 1 \pmod{2d}$ is a prime or a power of a prime and the code length parameter $n$ is a certain integer such that $n \leq (p-1)/(2d)$. The proposed code corrects double errors in the error set $\mathcal{E}_{p,d} = \{\pm 1, \pm \alpha^v, \ldots, \pm \alpha^{(d-1)v}\}$ where $\alpha$ is a primitive element of $GF(p)$ and $v = (p-1)/(2d)$.

The proposed code is suitable for the multidimensional signal constellations. For examples, it can be applied to correct at most double nearest neighbor errors on the square signal constellation in case of $d=2$ and on the hexagonal signal constellation in case of $d=3$. And specially, in case of $d=1$, the proposed code is equivalent to the double error correcting code, called negacyclic code proposed by Berlekamp [1] where a decoding algorithm for the negacyclic codes is proposed as well. After four decades, his decoding algorithm has been simplified by Roth [2], and recently Huber [3] has made Roth's algorithm much more accessible to readers in other fields.

The aim of this presentation is twofold. One is to construct double N-N error-correcting codes for the error set $\mathcal{E}_{p,d}$ mentioned above. The other is to give the corresponding decoding algorithm which corrects more general N-N errors than $\{+1, -1\}$ errors. And an example of the decoding process will be shown for $d = 3$, that is, N-N error correcting codes on a hexagonal signal constellation.

# On Error Detecting Codes

Yanling Chen and A. J. Han Vinck
Institute of Digital Signal Processing, University of Duisburg-Essen, Germany.
Email: {yanling.chen, han.vinck}@uni-due.de.

*Abstract*—This paper discusses the state-of-art of the error detecting codes. First, known methods are reviewed for detecting burst errors and human operation errors, respectively. Furthermore, we look into their connections and propose research challenges.

## I. Introduction

When information is transmitted over noisy communication systems, it may get scrambled by noise or data may get corrupted. To cope with this, error-detecting codes are often used to help us detect if an error has occurred during transmission of the message. Note that the system channel may be a telephone line, a high frequency radio link, or a satellite communication link. The noise may be caused by human errors, lightnings, thermal fluctuations, imperfection in equipment, etc. Due to different transmission medium and noise, errors may exhibit distinct statistical characteristics in various application scenarios.

To detect the errors, additional digits (i.e., check digits) are added to the data at the time of transmission. The data along with the check digits form a codeword. The whole philosophy of error detection is based on the assumption that the errors which a code is designed to detect are very frequent compared with the errors not guarded against by it. Therefore, to design a good error-detecting code, the very first and important task is to have a good understanding of the statistical behavior of the errors. In this paper, we will discuss two kinds of errors: 1) burst errors (that are common in many communication channels) and 2) human operation errors. Known error-detecting methods developed for both are reviewed, respectively.

## II. Known error-detecting methods

### A. Burst errors

Burst errors are common transmission errors in many communication channels, including magnetic and optical storage devices. To cope with such errors, a cyclic redundancy check (CRC) code [1] (based on polynomial division) is very suitable. Due to its easy implementation in binary hardware, numerous varieties of CRCs have been incorporated into technical standards.

### B. Human operation errors

Human operation errors could be divided into the following two classes: 1) Cognitive Errors and 2) Typographic Errors. For more detailed classification, one can refer to

This work is supported in part by DFG Grant CH 601/2-1.

the empirical investigations by Beckley [2] and Verhoeff [3].

There is a long list of research articles aiming to design good error-detecting codes against human operation errors, by using one of the following methods: a) division; b) parity check; c) weighted parity check; d) Luhn formula; e) hybrid system; f) Verhoeff's method; and g) polynomial. The underlying ideas include the basic decimal arithmetic (e.g., (a) division), make use of operation over groups [6] (e.g., (b) parity check and (c) weighted parity check methods), non-commutative operation over Dihedral groups (e.g.: (f) Verhoeff's method [3], [4]), and multiplication over finite fields (e.g., (g) polynomial method [5]) and vector-matrix multiplications in matrix algebra [7], [8].

## III. Open research problems

Generally speaking, a error detecting code can be described by $(n, t, \mathcal{A}, \mathbb{G})$, where $n, t$ are the lengths of the information digits and check digits, respectively; $\mathcal{A}$ is the alphabet (which size can be numerical, alphabetic, alphanumeric etc), and $\mathbb{G}$ is the operation group/quasigroup. Note that the well-developed error-control codes work over a finite field of a prime power order (and are analyzed for random errors). However, the error detection codes desired for specified applications may aim for detecting some kinds of atypical errors, and could have an alphabet size beyond this category. For instance, the best decimal code (with $|\mathcal{A}| = |\mathbb{G}| = 10$) is still Verhoeff's code [3] over the dihedral group $\mathbb{D}_5$, which was generalized by Gumm in [4] to a code with alphabet of size $2s$, $s$ odd.

# Extremality for Error Exponents of $q$-Ary Input Discrete Memoryless Channels

Yuta Sakai and Ken-ichi Iwata
University of Fukui, JAPAN, Email: {y-sakai, k-iwata}@u-fukui.ac.jp

*Abstract*—This study examines extremalities of arbitrary input discrete memoryless channels for Gallager's reliability function $E_0$ under a uniform input distribution.

## I. Introduction

In the context of channel coding theorems for block codes, relations among the probability of decoding error, the codeword length, and the coding rate are discussed. One of famous and classical results in that context is the error exponent, which characterizes an exponential error bound converging to either zero or one as the codeword length increases. Consider discrete memoryless channels (DMCs). Let random variables $X$ and $Y$ be the input and output of a DMC, respectively. We denote by $P_X$ and $W_{Y|X}$ the input and the transition probability distributions of a DMC, respectively. For a coding rate $R$ below capacity and a codeword length $N$, Gallager [4] gave an exponential upper bound on the probability of maximum likelihood (ML) error for good codes by deriving the random coding exponent[1]

$$E_r(R, P_X, W_{Y|X}) \coloneqq \max_{0 \le \rho \le 1} \Big( E_0(\rho, P_X, W_{Y|X}) - \rho R \Big), \quad (1)$$

where Gallager's reliability function $E_0$ is defined by

$$E_0(\rho, P_X, W_{Y|X}) \coloneqq -\log \sum_{y \in \mathcal{Y}} \left( \sum_{x \in \mathcal{X}} P_X(x) W_{Y|X}(y \mid x)^{\frac{1}{1+\rho}} \right)^{1+\rho}.$$

If $P_X$ is uniform, we simply write the $E_0$ function as $E_0(\rho, W_{Y|X})$. Later, Shannon et al. [12] and Arimoto [2] also derived the sphere packing and the strong converse exponents, respectively, as functions of the $E_0$ function to give exponential lower bounds on the probability of ML error for arbitrary codes.

In this paper, we investigate extremal $q$-ary input DMCs achieving the maximum and minimum values of $E_0(\rho, W_{Y|X})$ under a fixed another channel reliability function. Such extremalities were studied by Guillén i Fàbregas et al. [5] and Alsan [1] in binary-input cases, and by the authors in ternary-input cases. In the paper, we introduce a method for deriving extremalities of arbitrary input DMCs for $E_0(\rho, W_{Y|X})$ from sharp bounds on Rényi's information measures [10], [11].

This work was partially supported by JSPS KAKENHI, Grant-in-Aid for Scientific Research (C) 26420352 and 17K06422, and Grant-in-Aid for JSPS Research Fellow 17J11247.

[1]The error exponent $E_r$ is usually defined as $\max_{P_X} E_r(R, P_X, W_{Y|X})$, where the maximization is taken over all input distributions.

## II. Connection with Rényi's Information Measures

If $X$ follows a uniform distribution, then

$$E_0(\rho, W_{Y|X}) = H_\alpha(X) - H_\alpha(X \mid Y) \quad (2)$$

with $\alpha = 1/(1+\rho)$, where $H_\alpha(X)$ denotes the Rényi entropy [6] and $H_\alpha(X \mid Y)$ denotes Arimoto's conditional Rényi entropy [2]. In [10], [11], many sharp upper and lower bounds on Rényi's information measures were derived together with extremal probability distributions achieving the bounds. Combining these, we can establish extremalities of arbitrary input DMCs for $E_0(\rho, W_{Y|X})$. As a result, for example, extremal DMCs achieving the maximum and minimum values of the random coding exponents with a fixed different Rényi's information measure can be clarified. As summarized in [9, Section II], some notions of symmetry for DMCs and the method of channel symmetrization are useful to establish the extremalities for channel reliability functions under a uniform input distribution.

# Information Complexity

Ulrich Tamm
University of Applied Sciences
Interaktion 1, 33619 Bielefeld, Germany
Email: ulrich.tamm@fh-bielefeld.de

ABSTRACT

Communication complexity was introduced by A. Yao (1979) in the study of parallel computation. It found, since, surprising applications in many areas of computer science as VLSI layout, switching circuits, and auction theory.

A recent hot topic much related to communication complexity is information complexity. The setting can be described as follows.

Is it significantly easier to compute several instances of a problem simultaneously than sequentially? An affirmative answer to this question for communication complexity (the so-called direct-sum conjecture due to Wigderson) would have deep impact in circuit complexity. Recently, for probabilistic protocols the amortized (= simultaneous) communication complexity could be characterized as an information expression.

In the probabilistic setting, a randomized protocol can be regarded as a random variable $\Pi$, hence, a probability distribution on all possible communication protocols. Further, there are defined random variables $X$ and $Y$ on the inputs of the two communicators. The (internal) information complexity $I(\Pi, Y|X) + I(\Pi, X|Y)$ now is the sum of the mutual informations of the protocol with each of the communicators inputs.

It turns out that this information complexity characterizes the amortized communication complexity for probabilistic protocols. Such a single-letter characterization for some special functions was obtained for deterministic protocols by the Bielefeld group in the early 1990's. These results coincide with a general formula in this case conjectured in the survey by Braverman et al. in the IEEE Information Theory Society Newsletter.

# Feature Selection for Computer-Aided Detection of Cancer types

Peter H.N. de With, Fons van der Sommen, Patrick Langenhuizen and Joost van der Putten
Eindhoven University of Technology, PO Box 513, 5600 MB Eindhoven, NL

Barrett's Esophagus (BE) is a condition in which the body replaces the normal lining of the esophagus with an acid-resistant cell type, to prevent tissue damage caused by gastric reflux. This is correlated with an unhealthy lifestyle and it is a known precursor of esophageal cancer, yielding an over 30-fold increased chance of this deadly disease. For further developed cancer stages in brain tumors, cancer detection occurs frequently at later stage, but in such a case, cancer detection with imaging can support tumor control of vestibular schwannomas (VS) after gamma knife radiosurgery. Here, the detection during treatment can be exploited to correlate the cancer with pre-treatment growth rate to measure effectivity of the radiosurgery. A similar case occurs with pancreatic ductal adenocarcinoma (PDA), which is the fourth leading cause of cancer-related death with a 5-year survival rate in the western world. While surgery is currently the only potentially curative therapy, only 15-20% of the patients are considered for resection, since the rest of the patients have an advanced stage of cancer for which curative surgery treatment is too late. Here, preoperatively predicting the prognosis of PDA is important in the determination of the most appropriate treatment strategy.

In all of the above cases, cancer detection based on advanced image analysis is crucial to determine the characterization and behavior of the cancer to come to prediction of the treatment and/or resection decisions. The usefulness of computer-aided support for analysis of the disease has passed the point of discussion, but the reliable detection of specific forms of cancer still depends on applied types of imaging and the search for the best image features describing the cancerous tissue properties is still ongoing. In this paper we discuss the exploration of various image features to support a reliable and robust detection.

In the BE study, several popular computer vision methods for detection are evaluated. Furthermore, we introduce new, clinically-inspired features and compare the detection performance of the investigated methods to that of clinical endoscopic and VLE experts. Typical features for evaluation are spectral features like multi-directional Gabor wavelets and statistical features like the Gray-Level Co-occurrence Matrix (GLCM), Local Binary Patterns (LBP), Histogram of Oriented Gradients (HOG) features as well as features from the Fully Connected (FC) layers of ALexNet as used in Deep Learning. With a maximal Area Under the Curve (AUC) of 0.95 for the proposed features versus an AUC of 0.81 for the VLE experts, this study shows that computer-aided detection methods offer a quite promising technique for the interpretation of VLE data [1].

For VS detection in the brain, in an initial study Tumor Volume Doubling Time (VDT) prior to treatment was calculated using pretreatment MRI scan, at least 6 months before treatment, and the treatment MRI scan. Objective criteria of failure were applied, based on tumor volumes: after initial pseudo-progression period of 2 years, showing twice a volumetric tumor progression of at least 10%, consecutively. VDTs were correlated with observed tumor control rates and volumetric responses.

For PDA detection, we present the first experiments for automated determination of resectability using CT scans. The tumor features are extracted from a group of patients with both hypo- and iso-attenuating tumors of which 33 were resectable and 37 were not. The tumor contours are supplied by a medical expert. We present an approach that uses intensity, shape, and texture features to determine tumor resectability. A broad set of statistical features is compared and various feature selection algorithms are exploited to find the best descriptors. It was found that Logistic Regression of the feature vector containing a broad set of statistical features was the best to be combined with a decision tree algorithm, in order to obtain the highest classification result. Compared to expert predictions made on a similar data set, our method reaches better classification results, but the data set is still limited. We obtain up to 28% gain on correctly predicting non-resectability, which is essential for patient treatment.

# On Compression for Time Series Databases

Ashot N. Harutyunyan, Arnak V. Poghosyan and Naira M. Grigoryan
VMware
48 Mamikonyants str., Yerevan 0051, Armenia
Email: {aharutyunyan;apoghosyan;ngrigoryan}@vmware.com

Today's data centers or cloud computing infrastructures represent highly complex IT systems [1]. Reliable management of those systems requires monitoring [2] their performance or capacity indicators (millions of metrics) as time series data stored in big databases. It also requires applying data science and machine learning approaches to those data sets and to event data of the cloud infrastructure, including logs, for various management purposes and predictive tasks (anomaly and pattern detection, problem root cause determination, etc., see [3]–[6], where, particularly, information-theoretic methods can play an important role; see also notes [7] on information theory and big data problems). Measuring such a massive volume of data causes a high storage consumption problem. For archiving months of data from hundreds of thousands of compute/storage/network resources with hundreds of metrics each, terabytes of space is needed. The problem becomes critical for those users who plan to store years of data and one of options is quantization of individual time series subject to a fidelity criterion. However, in general, such a lossy compression must preserve the ?useful? patterns of the data. Another data reduction approach (cross-metric) we consider relies on the correlation content of the time series database employing *independent and principal component analysis*. It experimentally results in significant compression rates for real databases from IT environments.

# Graph based Satellite Image Time Series Analysis

Ferdaous Chaabane
GRESCOM Lab., Ecole Supérieure des Communications de Tunis, University of Carthage
SUP'COM, Cité technologique des communications, 2088 El Ghazala, Ariana, Tunisia
Email: Ferdaous.chaabane@supcom.tn

## Abstract

Scenes analysis through satellite images has always been a hot topic in the field of remote sensing. When the images are acquired at different dates for the same geographical area, they form a time series of satellite images known as Satellite Image Time Series (SITS). The temporal dimension combined with the spatial Very High Resolution (VHR), open a wide range of applications such as the study of land cover monitoring and classification according to occurring phenomena.

Primary, a description of STIS through their acquisition conditions and systems and their pretreatments are developed. To adopt a dynamic scene model, we opted for graphs whose nodes represent VHR cartographic objects and whose edges describe the temporal relationships.

The contributions of this work are discussed in two parts.

- First, a supervised classification using the SVM algorithm with a suited graph kernel is proposed and tested through an original modeling of nodes and arcs labels. This lead to classify the land plots temporal behavior and extract regions with stable evolution, periodic changes, growth, random behavior, etc.
- Second, we integrated the temporal dimension, through graphs modeling in a conceptualization approach of expert knowledge done for the study of a static scene. This approach allows introducing semantic for regions' evolutions from the point of view of a remote sensing expert by an ontology construction of a remote sensing scene and spatio-temporal knowledge base using concepts' histograms. The choice of the graph kernel as similarity measure has been proved through experimental results.

*Keywords:* Satellite Image Time Series (SITS), Very High spatial Resolution (VHR), dynamic scene, spatio-temporal evolution, SVM, graph kernel, expert knowledge, semantic remote sensing information, etc.

# Lossy Compression based on Binarized Autoencoder for Pattern Classification

Masayuki Imanishi and Tadashi Wadayama
Nagoya Institute of Technology, Japan
Email: wadayama@nitech.ac.jp

## I. INTRODUCTION

The advent of recent powerful neural networks (NN) triggered explosive spread of research activities and applications on deep neural networks (DNN). The DNN also have found many practical applications such as image recognition, speech recognition and robotics because of their excellent performance compared with traditional methods. For example, in image recognition competitions, it was recently shown that deep neural networks achieves recognition performance comparable to that of human. It is natural to exploit the potential of DNN for data compression applications. In this paper, we study a lossy encoding scheme based on binarized autoencoder. The proposed method is well matched to neural network-based pattern classifiers.

## II. AUTOENCODERS

Autoencoders are a well known class of neural networks. An autoencoder is trained so that squared Euclidean distances between inputs and outputs becomes small. Combined with a suitable regulation terms in the loss function, we can obtain a special representation of inputs at the hidden layer such as sparse representation of inputs.

The architecture of autoencoder with 3-layers are the following. Let $x \in \mathbb{R}^n$ be the input to the network. The values of the hidden layer are given by $h = f(Wx + b)$ where $W \in \mathbb{R}^{m \times n}$ is a weight matrix and $b \in \mathbb{R}^m$ is a bias vector. The function $f : \mathbb{R} \to \mathbb{R}$ is a nonlinear function called an *activation function*. The values of the output layers are given by $\hat{x} = \hat{f}(\hat{W}x + \hat{b})$ where $\hat{W} \in \mathbb{R}^{m \times n}$ and $\hat{b} \in \mathbb{R}^m$. The function $\hat{f} : \mathbb{R} \to \mathbb{R}$ is also an activation function. The set of parameters $\Theta = \{W, \hat{W}, b, \hat{b}\}$ are adjusted in a training phase so as to minimize the loss function $||x - \hat{x}||_2^2$ or another loss function.

## III. BINALIZED AUTOENCODER

The proposed lossy encoder consists of an autoencoder with a binary quantization hidden layer, which is called a *binalized autoencoder*. The values of hidden layer are quantized as $h = \text{sign}(Wx + b)$. The function $\text{sign}(a)$ gives value 1 if $a \geq 0$; otherwise it takes the value 0. We also introduce constraint $W = \hat{W}^T$. The vector $h$ can be considered as a compressed coded sequence corresponding to the input $x$.

## IV. COMPUTER EXPERIMENTS

We performed computer experiments for evaluating the compression performance of the proposed scheme. As a dataset, we used the MNIST handwritten number dataset in the experiments. A pixel data in the MNIST dataset has 784 pixels (28 × 28) and each white-black pixel has 8-bit depth. In this experiments, we used the sigmoid function $\sigma(a) = 1/(1 + \exp(-a))$ at the output layer, i.e., $\hat{f} = \sigma$. The loss function used in the training process is the cross entropy function defined by

$$-\frac{1}{nN_t} \sum_{i \in D_t} \sum_{j=0}^{n} \{x_j^{(i)} \log \hat{x}_j^{(i)} + (1 - x_j^{(i)}) \log(1 - \hat{x}_j^{(i)})\},$$

where $D_t$ is the $t$-th mini-batch and $N_t$ denotes the number of data items in the mini-batch $D_t$. In the experiments, we compressed randomly chosen pixel data from the MSNIT dataset and measured the compression rate and successful recognition rate. As a pattern classifier, a convolutional neural network with successful recognition rate 99.4% was used.

Figure 1 summarizes results of the computer experiments. The horizontal axis of Fig.1 represents the compression rate and the vertical axis indicates the successful recognition rate. The parameters used in the experiments are the following: $N_t = 100$, learning coefficient 0.25, the number of learning epochs 50000. The figure indicates that the proposed lossy encoder achieves successful recognition rate 97.8% with the compression rate 3.99%. As a benchmark, the results of run length coding (Wyle codes, compression rate 4.97%, successful recognition rate 99.5%) are also included in Fig.1. The details of the results will be reported at the workshop.

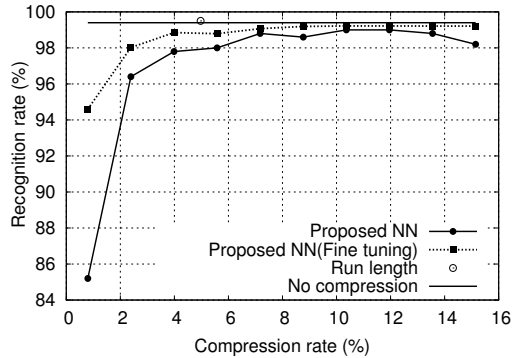

Fig. 1. Compression rate versus successful recognition rate

# The Information Bottleneck Method: Fundamental Idea and Algorithmic Implementations

Dirk Wübben
Department of Communications Engineering
University of Bremen, 28359 Bremen, Germany
Email: wuebben@ant.uni-bremen.de

*Abstract*—The quantized representation of signals is a general task of data processing. For lossy data compression the celebrated Rate-Distortion theory provides the compression rate in order to quantize a signal without exceeding a given distortion measure. Recently, with the Information Bottleneck method an alternative approach has been emerged in the field of machine learning and has been successfully applied for data processing. The fundamental idea is to include the original source into the problem setup when quantizing an observation variable and to use strictly information theoretic measures to design the quantizer. This paper introduces this framework and discusses algorithmic implementations for the quantizer design.

## I. Extended Abstract

A fundamental task in data processing is the quantized representation of noisy observations of an original source signal. Fig. 1 shows the considered system model consisting of a data source, a transmission channel and a quantizer.

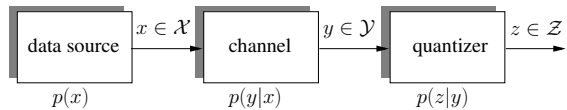

Fig. 1. General system model for the quantization of noisy observations

Without loss of generality, we assume the random variable x with realizations $x \in \mathcal{X}$ following the probability mass function (pmf) $p(x)$ as a discrete memoryless source (DMS). The observation variable y with realizations $y \in \mathcal{Y}$ is the output of a discrete memoryless channel (DMC) characterized by its transition probability distribution $p(y|x)$. Furthermore, the random variable z with realizations $z \in \mathcal{Z}$ is the output of the quantizer block being characterized by the conditional distribution $p(z|y)$. Thus, the quantizer output z is a compact representation of y with cardinality $|\mathcal{Z}| \leq |\mathcal{Y}|$. The design of the quantizer $p(z|y)$ realizes a trade-off between the *compression rate*, $I(y;z)$, and the *quality* of the compressed representation.

The Rate-Distortion (RD) theory provides the minimal number of bits per symbol in order to represent the received signal without exceeding an upper-bound on a given distortion measure, e.g., the mean square error (MSE) between the quantizer input signal and its representative at the output [1]. Specifically, the Blahut-Arimoto algorithm determines the lowest achievable compression rate for a certain maximum tolerable distortion. The main drawbacks of this formulation are the lack of a systematic way to choose a proper distortion measure for any case of pertinence and the fact, that the stochastic relation between the noisy observation and the original data source is not considered.

In [2], Tishby et al. have introduced the Information Bottleneck (IB) method for data compression. The central idea is to compress the observation y such that the quantizer output z preserves most of the information about the relevant variable, i.e., the original source x. Furthermore, IB avoids the a priori specification of a distortion measure by considering the mutual information $I(x;z)$ between the quantizer output and the original data source. In this fashion, the output of the quantizer becomes a compact representation of its input which is highly informative about the actual source of interest.

Given the joint probability distribution of the source and the channel output $p(x,y) = p(x)\,p(y|x)$ and assuming $x \leftrightarrow y \leftrightarrow z$ to be a Markov chain, the quantizer should be designed such that the output z is a compact representation of the input y which is highly informative about x. Mathematically, the existent trade-off between the *compression rate*, $I(y;z)$, and the *relevant information*, $I(x;z)$, is established by the introduction of a non-negative Lagrange multiplier, $\beta$, in the design formulation. Hence, for an allowed number of quantizer output levels, $n$, the corresponding design problem follows as [2]

$$p^\star(z|y) = \operatorname*{argmin}_{p(z|y)} \frac{1}{\beta+1}\bigl(I(y;z) - \beta I(x;z)\bigr) \quad \text{for } |\mathcal{Z}| \leq n \,. \quad (1)$$

The optimal quantizer mapping for (1) can be derived by means of variational calculus. Explicitly, for a specific value of $\beta$ the mapping $p(z|y)$ is a stationary point of the objective function in (1), if and only if

$$p(z|y) = \frac{p(z)}{\psi(y,\beta)} e^{-\beta D_{\mathrm{KL}}(p(x|y)\|p(x|z))} \quad (2)$$

is met for all pairs $(y,z) \in \mathcal{Y} \times \mathcal{Z}$. The function $\psi(y,\beta)$ normalizes the mapping $p(z|y)$ to ensure a valid distribution for each $y \in \mathcal{Y}$ and $D_{\mathrm{KL}}(\cdot\|\cdot)$ is the Kullback-Leibler (KL) divergence. The derived optimal mapping in (2) has an implicit form, as the cluster representative (in a conventional sense) $p(x|z)$ and the cluster probability $p(z)$ appearing on the right side of (2), depend on the quantizer mapping $p(z|y)$ by

$$p(z) = \sum_{y \in \mathcal{Y}} p(y)\,p(z|y) \quad (3)$$

and
$$p(x|z) = \frac{1}{p(z)} \sum_{y \in \mathcal{Y}} p(x,y)\, p(z|y) \ . \tag{4}$$

The iterative calculation of (2)-(4) leads to the Iterative Information Bottleneck (It-IB) algorithm [2]. Several alternative approaches to determine mapping functions in order to meet the trade-off between *compression rate* and *relevant information* have been discussed in the literature such as

- Agglomerative Information Bottleneck (Agg-IB) [3]
- Sequential Information Bottleneck (Seq-IB) [4]
- Deterministic Information Bottleneck (Det-IB) [5]
- KL-means Information Bottleneck (KL-means-IB) [6]
- Channel-Optimized Information Bottleneck (Ch-Opt-IB) [7].

For the special case of binary input alphabet computationally efficient adaptations exist as discussed in [8].

Subsequently, we compare the performance of the algorithms for 4-ASK ($x \in \{\pm 1, \pm 3\}$) input signals transmitted over AWGN channels with noise variance $\sigma_n^2 = 1$. Furthermore, to acquire the channel transition distribution $p(y|x)$, the continuous channel output is clipped at an amplitude of $3\sigma_n$ above the maximum input signal (i.e., 6 for 4-ASK) and uniformly discretized to $|\mathcal{Y}| = 128$ values. In particular, we investigate the accuracy by the *mutual information loss* $\Delta I = I(\mathsf{x};\mathsf{y}) - I(\mathsf{x};\mathsf{z})$ and the complexity-precision trade-off by the corresponding *compression rate* $I(\mathsf{y};\mathsf{z})$ for different values of $\beta$ over varying allowed number of clusters $n$.

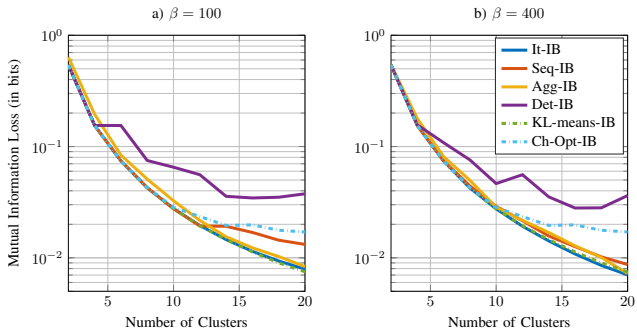

Fig. 2. Information loss $\Delta I$ for varying allowed number of bins $n$ and 4-ASK input alphabet with a) $\beta = 100$ and b) $\beta = 400$

Fig. 2 shows the information loss $\Delta I$ of the investigated algorithms. One may note, as the resultant mapping of all algorithms (except for the Agg-IB) depends on the initialization, to achieve the corresponding curves, they have been run $10^5$ times, with the best taken. Except for the KL-means-IB and the Ch-Opt-IB (both only consider $\beta \to \infty$) one can observe, that the accuracy of all algorithms is improved by increasing $\beta$ from 100 to 400. For a fair comparison with the KL-means-IB and the Ch-Opt-IB we concentrate subsequently on Fig. 2 b) with a relatively high value of $\beta$.

First of all, the non-smooth behavior of the Det-IB is due to the fact that its provided mapping does not necessarily use the entire allowed number of clusters, i.e., $|\mathcal{Z}| < n$. As an example, for $n = 12$ the used number of bins is smaller than the case of $n = 10$, leading to a coarser result. Furthermore, it can be seen that the It-IB and the KL-means-IB exhibit nearly the same performance over the entire range of allowed number of bins $n$. In addition, one notes that the Ch-Opt-IB also sweeps the corresponding curve of the It-IB for $n \leq 10$. The reason behind these observations is discussed in [9] where the asymptotic algorithmic equivalence of these algorithms is proven.

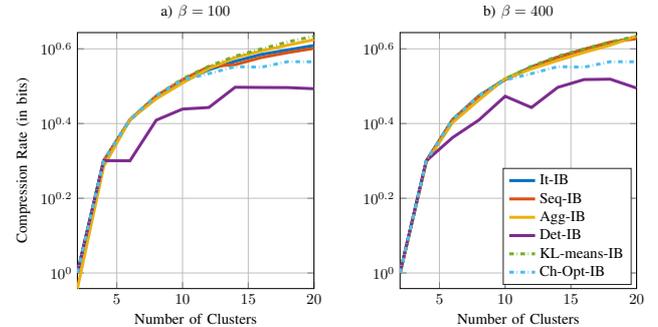

Fig. 3. Compression rate $I(\mathsf{y};\mathsf{z})$ for varying allowed number of bins $n$ and 4-ASK input alphabet with a) $\beta = 100$ and b) $\beta = 400$

Fig. 3 displays the corresponding compression rates $I(\mathsf{y};\mathsf{z})$. It can be observed, that in general, the lower the information loss introduced by quantization, the higher the corresponding compression rate.

ACKNOWLEDGMENT

This work was partly funded by the German ministry of education and research (BMBF) under grant 16KIS0720 (TACNET 4.0).

# The Max-LUT Method: Mutual-Information Maximizing Lookup Tables

Brian M. Kurkoski

Japan Advanced Institute of Science and Technology, Nomi, Japan

Email: kurkoski@jaist.ac.jp

The max-LUT method is an approach for implementing factor graph-based decoders and detectors using lookup tables (LUTs). In factor graphs, nodes compute output messages as a function of input messages. In the max-LUT method, these functions are implemented by LUTs, rather than by conventional mathematical operations. The LUT is designed to maximize mutual information between the output message and the information symbol it represents; this is a natural choice since maximizing mutual information leads to the highest achievable communications rates. The entire system is assumed to be discrete, since LUTs themselves produce discrete outputs from discrete inputs.

The max-LUT method is especially appealing for hardware implementations. For example, factor-graph LDPC decoders are commonly implemented in VLSI with fixed-precision arithmetic. Engineers often take a time-consuming trial-and-error approach to designing such decoders. The max-LUT method provides a systematic approach to design such decoders. In practice, such decoders have very good performance, when compared to other decoders of similar precision. Another example is implementation of computer simulations - the max-LUT method significantly reduces the simulation time of LDPC decoding algorithms.

Consider a general degree-three factor-graph node, Fig. 1. The factor graph node expresses a relationship f between three random variables $X_1, X_2$ and $X_3$, which are code symbols; this encoder-side perspective is shown in Fig. 1-(a). The factor graph node also has a decoder-side perspective, shown in Fig. 1-(b). Three random variables $L, Z$ and $V$ are messages about the code symbols $X_1, X_2$ and $X_3$, respectively, where $L$ and $Z$ are inputs and $V$ is the output. In the case of a binary LDPC code check node, the function $f$ expresses $f(X_1, X_2) = X_1 + X_2$ with modulo-two arithmetic. In the case of an binary LDPC variable node, the function $f$ expresses:

$$f(X_1, X_2) = \begin{cases} X_1 & \text{if } X_1 = X_2 \\ \text{undefined} & \text{if } X_1 \neq X_2 \end{cases}. \tag{1}$$

The conditional distributions $p_{L|X_1}(\ell|x_1)$ and $p_{Z|X_2}(z|x_2)$ are known. This may be obtained from the channel noise model, for example.

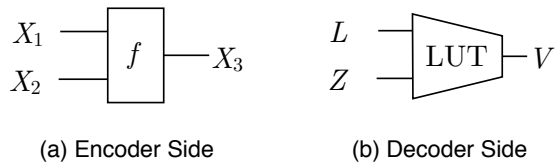

(a) Encoder Side  (b) Decoder Side

Fig. 1: (a) Encoder-side representation of a factor graph. (b) Decoder-side representation of factor graph.

With these inputs, the max-LUT method proceeds as follows, and is shown in Fig. 2. It is straightforward to construct the conditional distribution $p_{L,Z|X_3}(\ell, z|x_3)$ from the given conditional distributions and $f$. It is desired to find a function LUT which computes a new variable $v$ from the input variables $\ell, z$:

$$v = \text{LUT}(\ell, z). \tag{2}$$

To do this, find the LUT* which maximizes mutual information $I(V; X_3)$ :

$$\text{LUT}^* = \arg\max_{\text{LUT}} I(V; X_3) \text{ or} \tag{3}$$
$$= \arg\max_{\text{LUT}} I(\text{LUT}(L, Z); X_3). \tag{4}$$

This maximization problem is equivalent to designing a quantizer for a discrete memoryless channel, and is connected with the classification problem in machine learning. If the alphabet size of $L$ is $|\mathcal{L}|$ and the alphabet size of $Z$ is $|\mathcal{Z}|$,

This work was supported in part by JSPS Kakenhi Grant Number JP 15KK0005 and Grant Number JP 16H02345.

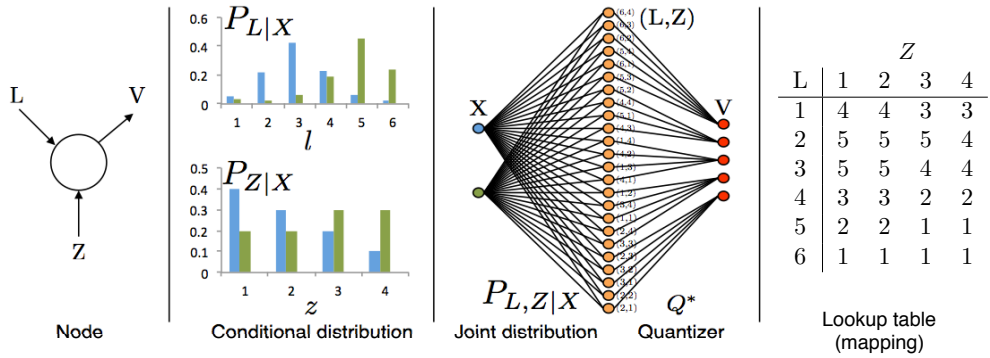

Fig. 2: Summary of the max-LUT method, generating a lookup table for a factor graph node.

then the discrete memoryless channel has $|\mathcal{L}| \cdot |\mathcal{Z}|$ outputs. Moreover, the mapping from $(\ell, z)$ to $v$ provided by the quantizer gives the lookup table LUT.

There are various approaches to find LUT. When the $X_i$ are binary, there is an optimal dynamic programming approach. When $X_i$ are non-binary, efficient but suboptimal approaches such as the information bottleneck method, or the KL-means algorithm (which is the K-means algorithm with Kullback-Leiber divergence metric).

The above method has been applied to decoding of LDPC codes. In a typical case the size of the alphabets $|\mathcal{L}| = |\mathcal{Z}| = |\mathcal{V}| = 16$ (corresponding to 4 bits/message), gives performance that is similar to floating point messages on the factor graph.

# An Information Bottleneck LDPC Decoder on a DSP

Jan Lewandowsky, e-mail: jan.lewandowsky@tuhh.de

Hamburg University of Technology, Institute of Communications, 21073 Hamburg, Germany

**Scope and Motivation:** Recently, *discrete* message passing decoders for low-density parity-check (LDPC) codes have attracted some attention [1, 2]. Discrete LDPC decoders are designed using mutual information maximizing clustering algorithms, such as the Information Bottleneck method, in a discrete density evolution scheme. The clustering algorithms are applied to obtain lookup tables for all decoder iterations. These lookup tables replace the classical check node and variable node operations in the decoder. Software defined radio (SDR) platforms play an important role in modern communications because they offer very high flexibility. On an SDR platform, most parts of the signal processing chain are implemented in software that is run on a digital signal processor (DSP), a general purpose processor (GPP) or heterogeneous hardware. The aim of the presented work was to validate the assumption that the integer based decoding process in discrete decoders which results from purely information theoretical considerations also offers *practical* gains in a decoder implementation suitable for SDR applications. For this purpose, a discrete decoder was implemented on a Texas Instruments TMS3206474 DSP using a TMDSEVM6474L evaluation board and compared with several state-of-the-art message passing decoders on the same hardware.

**Challenges:** Naturally, memory addressing and memory access latency have strong influence on the decoding throughput of discrete decoders. The DSP natively supports standard 8, 16 and 32 bit data types. However, discrete LDPC decoders show very good performance also with shorter messages. A bit width of 4 bit is sufficient to perform very close to double precision belief propagation decoders for binary input additive white Gaussian noise (AWGN) channels. Embedding 4 bit data in standard data types requires additional operations for bitfield extraction. It was a challenge to find a good trade-off between memory usage of the decoder and fast memory access. We developed a technique which accepts larger lookup tables for a simpler addressing operation of the lookup table entries in the decoder. Moreover, we studied several methods to obtain a decoder implementation with a small memory footprint. The lookup tables holding the decoding mappings for all decoder iterations are accessed extremely often during the iterative decoding process. The TMS320C6474 DSP is not especially optimized for fast memory access. However, we found it to be beneficial for the discrete LDPC decoding process that the DSP has the huge number of 32 registers. It was a challenge to develop a discrete decoder implementation which makes use of this strength of the selected hardware platform.

**Results:** The novel discrete LDPC decoder was tested for a quantized output AWGN channel with binary phase shift keying modulation and exemplary $(3,6)$-regular LDPC codes for varying $E_b/N_0$. The maximum number of decoder iterations was 50. The decoder stopped decoding earlier, if all parity checks were satisfied. Messages occupying 4 bit were passed for the decoding in the discrete decoder. Figure 1 compares the achieved decoding throughput for a length 8000 LDPC code on the DSP. The discrete decoder was found to offer higher throughput and also better bit error rate performance than the well known min-sum decoder in our implementation. Moreover, its throughput exceeds the throughput of a min-sum decoder which uses a lookup table holding the jacobian logarithm correction function, but the discrete decoder offers almost identical bit error rate performance. For the results shown in Figure 1, the implementations of the conventional decoders used a Q8.4 fixed-point implementation.

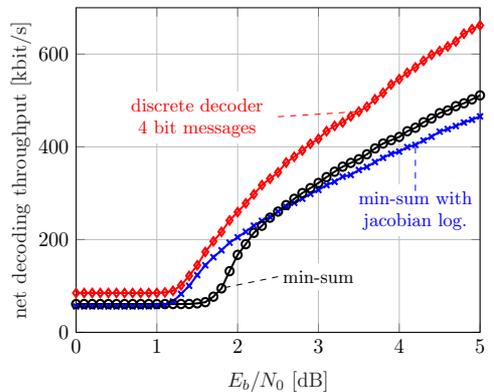

Figure 1: Decoding throughput of several iterative decoders on the TMS320C6474 DSP.

**Acknowledgment:** Special thanks to Peter Oppermann who did all decoder implementations in the scope of his master thesis, to Professor Gerhard Bauch for supervision of the thesis and to Matthias Tschauner, Marc Adrat and Markus Antweiler from the Fraunhofer Institute for Communication, Information Processing and Ergonomics FKIE in 53343 Wachtberg, Germany for their highly appreciated support of this work.

# AIFV Codes and Related Topics

Hirosuke Yamamoto
Dept. of Complexity Science and Engineering, The University of Tokyo
Kashiwa-shi, Chiba, 277-8561, Japan. Email: Hirosuke@ieee.org

It is well known from Kraft's and McMillan's theorems that the Huffman code, which is optimal in the class of instantaneous codes (i.e. prefix codes), is also optimal in the class of uniquely decodable codes if a fixed single code tree is used. However, AIFV (almost instantaneous fixed-to-variable length) codes were proposed recently [1], [2], which uses two (resp. $K-1$) code trees for the binary (resp. $K$-ary) code alphabet, and it was shown that the AIFV codes can attain better compression rate than the Huffman codes for i.i.d. sources. Since source symbols are assigned to incomplete internal nodes in addition to leaves in the AIFV codes, they are not instantaneous codes. But, they are devised so that the decoding delay is at most 2 bits (resp. one code symbol) for the binary (resp. $K$-ary) codes. Hence the codes are called *almost instantaneous*.

The average code length of the binary Huffman code $L_{\text{H}}$ is bounded by $H(X) \leq L_{\text{H}} < H(X)+1$, and the upper bound is tight when $p_{\max} \to 1$, where $p_{\max}$ is the probability of the most probable source symbol, i.e. $p_{\max} = \max_{x \in \mathcal{X}} p(x)$. On the other hand, the average code length of the optimal binary AIFV code $L_{\text{AIFV}}$ is bounded by $H(X) \leq L_{\text{AIFV}} < H(X)+0.5$ and the upper bound is tight when $p_{\max} \to 1$ [3]. Note that $L_{\text{H}}$ cannot become shorter than one in any case, but $L_{\text{AIFV}}$ can become shorter than one when $p_{\max}$ is large.

The binary AIFV codes, which use two code trees and have at most 2-bit decoding delay, can be extended to the binary AIFV-$m$ codes, which use $m$ code trees and have at most $m$-bit decoding delay for $m \geq 2$. The binary AIFV-2 codes correspond to the original binary AIFV codes. Furthermore, it is proved that the average length of the optimal AIFV-$m$ code $L_{\text{AIFV-}m}$ is bounded by $H(X) \leq L_{\text{AIFV-}m} < H(X)+\frac{1}{m}$ for $2 \leq m \leq 4$ [3]. We note that the Huffman codes for extended alphabet $\mathcal{X}^m$ can also attain the same bound. But, in the case of Huffman codes for $\mathcal{X}^m$, the size of the code tree is $O(|\mathcal{X}|^m)$ and the largest decoding delay becomes much larger than $m$ bits for the first symbol of $x^m \in \mathcal{X}^m$ since $m$ source symbols $x^m$ are decoded at once. In contrast, in the case of the AIFV-$m$ codes, the total size of code trees is $O(m|\mathcal{X}|)$ and the decoding delay is at most $m$ bits. Hence, the AIFV-$m$ codes are superior to the Huffman codes for $\mathcal{X}^m$ in terms of memory size and decoding delay.

For a given source, the optimal AIFV code can be constructed by an iterative optimization algorithm combined with integer programing [2] or dynamic programming [4]. Furthermore, the optimal binary AIFV-$m$ code trees can also be constructed by the iterative optimization algorithm combined with a dynamic programming algorithm of time complexity $O(mn^{2m+1})$ and space complexity $O(mn^{m+1})$ for $2 \leq m \leq 5$, where $n = |\mathcal{X}|$ [5].

For sources with $n = |\mathcal{X}|$, the total number of binary Huffman code trees is given by the $(n-1)$-th Catalan number $C_{n-1}$ while the total number of binary AIFV code trees is given by the $(n-1)$-th large Schröder number $S_{n-1}$, which is the total number of Schröder paths [6]. Since it holds that $\log_2 C_n = 2n + o(n)$ and $\log_2 S_n = n\log_2(3+2\sqrt{2}) + o(n) \approx 2.5431n + o(n)$, binary AIFV codes have a great variety of code trees compared with Huffman codes. Since there is a bijection between the set of binary AIFV code trees and the set of Schröder paths, each AIFV code tree can be represented with $\lceil \log_2 S_{n-1} \rceil$ bits by using the bijection [6].

If the probability distribution of a source is unknown, we can use the adaptive (dynamic) Huffman code instead of the static Huffman code. In the case of AIFV codes, it is very difficult to obtain an algorithm to update code trees adaptively and optimally for given source sequences. However, it is shown that (non-optimal) binary AIFV code trees can easily be constructed from the Huffman code tree by applying a certain operation to each pair of sibling with a leaf such that the probability weight of the leaf is more than twice as large as the one of the other sibling node [3]. Hence, adaptive AIFV coding can be realized by the AIFV code trees derived from the Huffman code tree which is updated adaptively by the dynamic Huffman algorithm. For source sequences with $p_{\max} > 0.5$, the above adaptive AIFV code can improve compression rate considerably compared with the dynamic Huffman code [7].

# On the Characterization and the Modelling of the Narrowband PLC Channel Transfer Function and Noise

Hela Gassara, Fatma Rouissi and Adel Ghazel
GRESCOM Lab., Ecole Supérieure des Communications de Tunis, University of Carthage
SUP'COM, Cité technologique des communications, 2088 El Ghazala, Ariana, Tunisia
Email: {hela.gassara, fatma.rouissi, adel.ghazel}@supcom.tn

ABSTRACT

Power Line Communication (PLC) has always been considered as a promising technology for Smart Grid applications such as advanced metering infrastructure (AMI), demand response (DR), home automation, in-home environment and vehicle-to-grid communications. Many narrowband PLC systems have appeared operating either in the CENELEC bands (3-148.5 kHz) in Europe or in the FCC band (9-490 kHz) in USA or in the ARIB band (9-450 kHz) in Japan, and delivering up to 500 kbps.

The design of a coupling interface between the power line and the measuring instruments with conformity to immunity requirements against electromagnetic disturbances allowed to build an experimental set-up for measuring the characteristics of the indoor low-voltage electrical network in ENELEC/FCC/ARIB bands. Obtained measurements results were exploited for conducting the statistical characterization and the deep study of the time-frequency characteristics of the low-frequency PLC channel, especially the path loss, the average channel gain, the delay spread parameters, the coherence bandwidth and the channel capacity.

Following the top-down modelling approach, a new parametric model was proposed for the narrowband PLC channel transfer function taking into account power lines transmission characteristics at the low frequency range. The proposed model was validated through field measurements in the CENELEC/FCC/ARIB bands using the Matrix Pencil algorithm for model parameters identification. The statistical distributions of the model parameters were then determined allowing the random generation of narrowband PLC channels which are in good agreement with the experimental ones.

The power spectral density of the stationary noise, composed of background noise and narrowband interferences, was experimentally characterized and statistically modelled in the frequency band from 9 to 500 kHz. The model has a limited number of parameters which were considered as random variables approximated by their corresponding statistical distributions. A stochastic model was also established for the impulsive noise present in the indoor narrowband PLC environment. An appropriate method was used to finely extract pulses and their characteristic parameters based on extensive measurements carried out inside several buildings. These pulses, with pseudo-frequency below 500 kHz, were statistically modelled by fitting parameters deduced from measurements with proper distribution functions. The generation of the noise present on the power line was then preceded at CENELEC/FCC/ARIB bands.

The established stochastic models of channel transfer function and noise are useful for assessing the performance of different communication chains operating in frequencies below 500 kHz and for improving the design of narrowband PLC systems.

# Low Complexity and Reliable Narrow Band PLC Modem for Massive IoT Nodes

Safa Najjar*, Fatma Rouissi*, Adel Ghazel* and A.J. Han Vinck†
* GRESCOM Lab., Ecole Supérieure des Communications de Tunis, University of Carthage
SUP'COM, Cité technologique des communications, 2088 El Ghazala, Ariana, Tunisia
† Institute of Digital Communications, Duisburg-Essen University, Germany
Email: {Safa.najjar, fatma.rouissi, adel.ghazel}@supcom.tn; han.vinck@uni-due.de

ABSTRACT

This research presentation will address the design challenges to meet the low-cost and sufficient quality of service requirements for Narrow Band Power Line Communication (NB-PLC) Modem for massive IoT applications where the existence of power supply cables can make PLC more competitive than wireless. The target applications cover mainly Indoor low bite rate communication services for home automation and building control. Important industrial development has been carried out to produce and deploy NB-PLC in the low frequency band less than 500 kHz by using single carrier modulations. Multiple standards BACnet [1], KNX (ISO/IEC 14543-3-5, EN 50090) [2], and LON (ISO/IEC 14908-3, ANSI 709.2) [3], [4] have been released and are being considered for future Smart Home services. This category of NB-PLC is not succeeding to reach the great commercial success through mass deployment mainly due to the strong competition of low-cost short range wireless solutions.

Today, the massive IoT deployment by using wireless technology is facing the big challenge of limited radio spectrum resources and more restrictive EMC regulation that is limiting the indoor radio propagation inside buildings. This offers an important opportunity to promote NB-PLC that will allow relaxing the occupation of the radio spectrum and overcome indoor poor radio propagation issues. Following the top-down modelling approach, a new parametric model was proposed for the narrowband PLC channel transfer function taking into account power lines transmission characteristics at the low frequency range. The proposed model was validated through field measurements in the CENELEC/FCC/ARIB bands using the Matrix Pencil algorithm for model parameters identification. The statistical distributions of the model parameters were then determined allowing the random generation of narrowband PLC channels which are in good agreement with the experimental ones.

This observation is motivating our research work within the framework of a Tunisian-German scientific collaboration project to propose an optimized design of a NB-PLC Modem by exploring the efficient and robust single carrier modulation combined with a low-complexity channel coding scheme. In order to adapt the modulation technique to the target indoor powerline environment in the CENELEC (9-490 kHz), FCC (9-490 kHz) and ARIB (9-450 kHz) bands, the PLC channel frequency response in the presence of background and impulse noise are considered to compare modulation techniques and define the optimal demodulation processing architecture [5], [6]. Obtained simulation results, by using measured indoor channel frequency response and noise, are showing superior performances with M-ary FSK combined with permutation code [7]. Considering the impact of the PLC Modem hardware implementation on the communication performances and the product cost this research work is also studying the cost effective optimal performance design. Special interest is given to the robustness analysis of the coupling and analog front-end interface to overcome the power line impedance variation with a standard compliant design. Cost effective objective is explored through a fully digital Modem implantation giving the ability to adjust future parameters to suit PLC network needs.

This talk will present the intermediate research results relative to the PLC narrow-band low-complexity communication scheme for the modulation technique choice and the adapted channel coding for the PLC channel attenuation and impulsive noise. Ongoing studies relative to the implementation design will be exposed and discussed by considering the state of the arts of market products.

# CFAR Algorithm based Millimetric Radar Detection Enhancement for ADAS Applications

Khaled Grati*, Hamed Gharbi†, Fatma Rouissi* and Adel Ghazel*
* GRESCOM Lab., Ecole Supérieure des Communications de Tunis, University of Carthage
SUP'COM, Cité technologique des communications, 2088 El Ghazala, Ariana, Tunisia
† EBSYS, Wevioo Group, Technopôle Elgazala, 2088 El Ghazala, Ariana, Tunisia
Email: {Safa.najjar, fatma.rouissi, adel.ghazel}@supcom.tn; hamed.gharbi@wevioo.com

ABSTRACT

This presentation deals with ADAS (Automatic Driver Assistance System) Radar sensor that is responsible for the detection of potential collisions or hazardous situations to warn/alert the driver or to intervene with the braking and other controls of the vehicle in order to prevent an accident. The ADAS Radar system is characterized by the key performance parameters: Detection range, Speed detection Range, Range precision, Velocity precision, Angular resolution and Angular width of view.

By considering short-range and mid-range ADAS radar (range of tens of meters) applications, such as blind-spot detection, pedestrian detection and pre-crash alerts, algorithmic enhancement are for the range and the speed accuracy. The radar transmission waveform and the receiver processing techniques should not only be able to detect targets with high probability but also should automatically maintain the false alarm rate at a constant level by adaptively adjusting the detection threshold according to the background clutter and noise using a constant false alarm rate (CFAR) detector.

Figure 1 represents the detection processing stages for an FM-CW (Frequency Modulation Continuous Wave) with saw tooth transmission waveform.

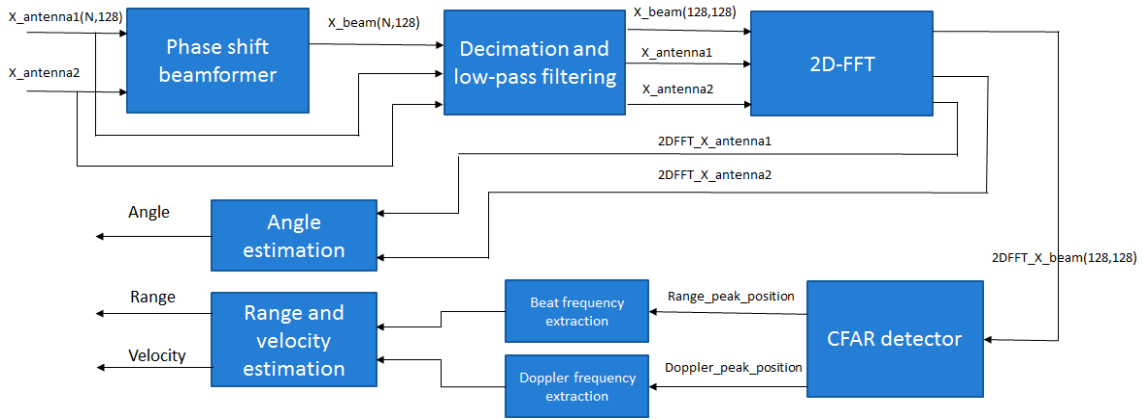

Fig. 1: Radar detection processing stages for FM?CW saw tooth transmission waveform.

The proposed radar signal processing is implemented on the Infineon TC26xB 23-bit microcontroller. To generate the waveform signal, Infineon 24 GHz MMIC 1TX/2RX transceiver is used with an external SPI DAC to control the output frequency. Data transfer between DAC, ADC and FFT module is controlled by the DMA Channels. The data sampling and the first stage FFT are executed in parallel.

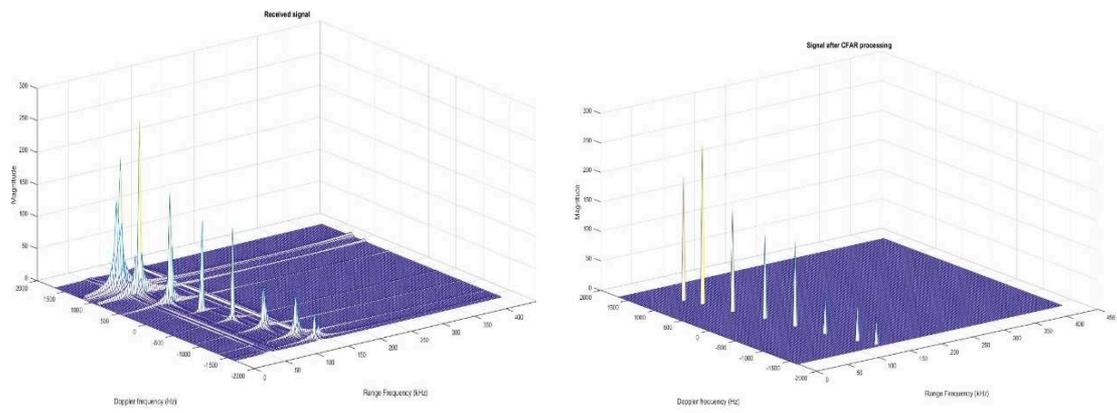

Fig. 2: Figure 2. 24 GHz Radar detection results (a) before CFAR processing; (b) after CFAR processing .

# On Universal FV Coding Allowing Non-vanishing Error Probability

Shigeaki Kuzuoka
Faculty of Systems Engineering
Wakayama University
Sakaedani 930, Wakayama-shi, Wakayama, 640-8510 Japan
Email: kuzuoka@ieee.org

## I. Summary

In this talk, we revisit a *folklore* which is stated as follows:[1]

> Given a *good* variable-length lossless code with codeword length $\ell_n(x^n)$ for each $x^n \in \mathcal{X}^n$, the set $\mathcal{C}_n = \{x^n : \ell_n(x^n) \leq nR\}$ of sequences for which the codeword length $\ell_n(x^n)$ is smaller than or equal to the given threshold $nR$ gives a *good* fixed-length code with the coding rate $R$.

In the statement above, the *goodness* of codes is not well defined. The purpose of this study is to give precise formulation for this folklore. In particular, we consider this problem from the aspect of *universal source coding* [2], [3].

In summary, we apply universal fixed-to-variable length (FV) lossless codes to
1. fixed-to-fixed length (FF) coding allowing non-vanishing error probability, as known as $\varepsilon$-FF coding, and
2. FV coding allowing non-vanishing error probability, as known as $\varepsilon$-FV coding.

Our results demonstrate that we can construct a universal $\varepsilon$-FF code and a universal $\varepsilon$-FV code from a universal FV lossless code.[2]

## II. Key idea ("Concept" of this talk)

The results of this talk depend heavily on the result of Koga and Yamamoto [6, Theorem 2]: Assume that the codeword length $\ell_n(x^n)$ of an FV lossless code satisfies,[3] for an information source $\boldsymbol{X} = \{X^n\}_{n=1}^{\infty}$,

$$\frac{1}{n}\mathbb{E}\left[\ell_n(X^n)\right] - \frac{1}{n}H(X^n) \to 0 \quad (n \to \infty),$$

where $H(X^n)$ denotes the entropy of $X^n$. Then, under a certain condition,

$$\frac{1}{n}\ell_n(X^n) - \frac{1}{n}\log\frac{1}{P_{X^n}(X^n)} \to 0$$

in probability as $n \to \infty$, where $P_{X^n}$ denotes the probability distribution of $X^n$.

Roughly speaking, the above result guarantees that we can approximate the *entropy-spectrum* of the source (i.e., the distribution of the *entropy density rate* $(1/n)\log(1/P_{X^n}(X^n))$) by using the distribution of the codeword length per symbol $\ell_n(X^n)/n$.

Based on this fact, we can demonstrate the following results:

(1) If there exists a universal FV lossless code for a set of sources $\Omega$ then the set $\mathcal{C}_n = \{x^n : \ell_n(x^n) \leq nR\}$ of sequences for which the codeword length $\ell_n(x^n)$ is smaller than or equal to the given threshold $nR$ gives a universal FF code allowing errors for $\Omega$.

(2) Let us construct an FV code with codeword length $\hat{\ell}_n(x^n)$ such that
  (i)
  $$\hat{\ell}_n(x^n) = \begin{cases} 1 + \ell_n(x^n) & \text{if } x^n \in \mathcal{C}_n \\ 1 & \text{otherwise.} \end{cases}$$

  (ii) All $x^n \in \mathcal{C}_n$ can be reconstructed at the decoder.

Then, the constructed FV code can achieves the optimal trade-off between the average codeword length and the error probability universally for $\Omega$.

## Acknowledgment

This work was supported by JSPS KAKENHI Grant Number 26820145.

---

[1] The connection between the error probability of FF codes and the overflow probability of FV codes, which is shown by Uchida and Han [1], gives a theoretical support for this folklore.

[2] A similar problem was studied by the author [4], [5], where relations between universal FV coding and FF coding with vanishing error probability are investigated. In this sense, this paper gives a slight generalization of [4], [5].

[3] It may be worth to note that the codeword length of an FV lossless code can be characterized by an integer-valued function $\ell_n(x^n)$ satisfying the Kraft inequality $\sum_{x^n} 2^{-\ell_n(x^n)} \leq 1$.

# On Prediction using a Data Base: An Information-Theoretic Approach

Stan Baggen

Philips Lighting Research, High Tech Campus 7, 5656 AE Eindhoven, The Netherlands

e-mail: stan.baggen@philips.com

EXTENDED ABSTRACT

*Introduction*

In data science, one wants to extract relevant information from a possibly large data base. It turns out that we may indeed obtain useful results using information-theoretic methods, if we are able to define an operational means of what is relevant.

An example could be a data base of a machine $M$, which contains the "class" $y$ of $M$ as a function of time $t$, where in the remainder the class $y \in \{0, 1\}$ is assumed to be binary, and it corresponds to $M$ being "up" or "down". Furthermore, the data base contains the recordings of many sensors $\bar{x}$ of $M$ (often called "features" in data science), also as a function of time $t$. For notational convenience, we henceforth use the notation $x$ for the random variables corresponding to the recording of the features, both for the scalar and the vector case. The relevant information to be extracted from the data base can now, for example, be defined as finding a possibly simple predictive model that is able to predict reliably the class $y$ of machine $M$ at some point in the future, given the recorded realizations of all sensors $x$ in the past up to the current time $t_0$. Mathematically, one is looking in this case for a possibly simple $p(y|x)$, which represents a stochastic function of how $y$ depends on the features $x$.

It turns out that very often, most of the information in the data base is of little or no importance in establishing such a model. A major part of the time of a data analist is devoted to establishing which components of $x$ might be relevant for predicting $y$, and how to quantize and combine these components in a useful, relatively simple model.

Information-theoretic methods, as explained hereafter, may actually formalize these procedures in an optimal and automated manner. As mentioned before, these procedures should be able to discard most of the available information in the data base, e.g.,

- non-relevant information,
- too much detail,
- noise,
- errors,

in such a manner, that the relevant information is presented in a concise and useful model. As we explain in the remainder, such discarding while retaining the relevant information, can be seen as a form of quantization of the data base.

In a seminal article, Tishby et al. propose the Information Bottleneck Method [1]. In this paper, the authors present an information-theoretic way of quantifying the concept "retaining relevant information" in a useful manner. Their way of thinking can be seen as an extension of rate distortion theory.

Classical rate distortion theory sets information-theoretic bounds on what can be achieved by (vector) quantizers in terms of a trade-off between distortion of a source signal and the bit rate for representing the source [2], [3]. Also in the Information Bottleneck method, retaining of relevant information is approached by assuming that such can be achieved by a proper quantization of the data base, which we from now on call "Tishby quantization".

In rate distortion theory, a rate distortion function $R(D)$ is introduced, which provides a bound on the performance of (vector) quantizers for a given source, in terms of a minimal average distortion $D$ that is attainable for a certain average bit rate $R$. Let's assume that a fixed vector quantizer $Q$ maps the signal $x$ onto a quantized version $x'$, $Q : x \mapsto x'$. In classical rate distortion theory, known distortion measures are e.g.:

- Euclidean distance: $d_E(x, x') = ||x - x'||^2$,
- Hamming distance: $d_H(x, x') = \text{wt}(x - x')$, where $\text{wt}(z)$ corresponds to the number on non-zero components of $z$.

In classical rate distortion theory, everything revolves around the variables $x$, its quantized version $x'$, and a distortion that only depends on $x$ and $x'$.

In the setting of predicting $y$ using a set of quantized features $x'$, i.e., generating a "good" $p(y|x')$, the idea put forward by Tishby et al., is to define a Kullback-Leibler (KL) distortion measure on $p(y|x')$, where now the random variable $y$ also is included. Tishby et al. define

$$d_{KL}^{(y)}(x, x') = \sum_y p(y|x) \log \frac{p(y|x)}{p(y|x')}, \qquad (1)$$

where the distortion $d_{KL}^{(y)}(x, x')$ reflects the loss of relevant information concerning $y$ when features $x'$ are used for predicting $y$ in stead of using features $x$. As mentioned before, one may think of $x'$ as being (vector) quantized versions of $x$.

In Fig. 1, we depict an example of the idea of Tishby quantization in a block diagram. In the example, the subject of interest is a machine $M$ that is characterized by a class $y$ ("up" or "down") as a function of time. Furthermore, $M$ outputs a set of features $x$ as a function of time. We have not indicated the time $t$ explicitly, as we may be interested in the information that past realizations of $x$ may provide about the current $y$. Moreover, different components of the vector $x$ may refer to the same physical sensor taken at different times in the past.

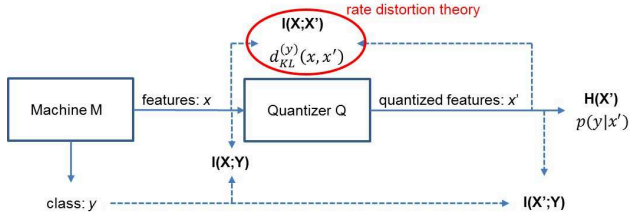

Fig. 1: Example of Tishby Quantization

The mutual information $I(X;Y)$ corresponds to the average amount of relevant information that realizations of the features $x$ may provide about the class $y$. If we quantize features $x$ into $x'$, we incur a KL-distortion $d_{KL}^{(y)}(x,x')$ as given by (1). According to rate distortion theory [3], there exist quantizers $Q : x \mapsto x'$ that on average only need a rate $R(D)$ bits per transmission for representing the signal $x$ with an average KL-distortion at most $D$, where

$$R(D) = \min_{p(x'|x):\sum_{x,x'} p(x)p(x'|x)\cdot d_{KL}^{(y)}(x,x')\leq D} I(X;X'). \quad (2)$$

If the quantizer is deterministic, we have in addition

$$H(X'|X) = 0 \quad \Rightarrow \quad I(X;X') = H(X'). \quad (3)$$

As a result from Tishby quantization, we finally have an efficient description $x'$ of new features that can be extracted from the data base, costing on average $H(X')$ bits per realization for transmission (or storage), with the corresponding $p(y|x')$ for predicting the class $y$ using $x'$, which still provides the largest possible amount of information $I(X';Y)$ concerning the class $y$ of machine $M$ (given that we started with features $x$).

*Kullback-Leibler Distortion*

The Kullback-Leibler distortion (1) or information divergence [3], [4] between the conditional distributions $p(y|x)$ and $p(y|x')$ is a measure for the amount of bits of information that is lost, if one uses $x'$ to predict $y$ in stead of $x$. One can show easily that

$$E_{x,x'}\left[d_{KL}^{(y)}(x,x')\right] = I(X;Y) - I(X';Y), \quad (4)$$

hence, the average KL-distortion between $p(y|x)$ and $p(y|x')$ equals the loss in average mutual information about $Y$ due to the quantization of $x$ into $x'$, where $Y$ is the relevant quantity of the data base.

*Modified Lloyd Algorithm*

Tishby et al. [1] discuss information-theoretic consequences of their approach, in particular of finding bounds that correspond to a rate-distortion function $R(D)$. For finding (an approximation to) a Tishby quantizer of a realization of an unknown data source, one may consider a modification of the well-known extended Lloyd algorithm [5], as is also discussed by [6], [7]. The modified Lloyd algorithm is an iterative algorithm, containing three steps in each iteration, which also allows for estimating the cost of transmission:

1) given a partitioning of $x$, find the best representative $x'$ in each partition:

$$x' = \operatorname{argmin}_{x''} E_{x\in\text{partition}}\left[d_{KL}^{(y)}(x,x'')\right] \quad (5)$$

2) given a partitioning of $x$ and the resulting $p(x')$ on the representatives of the partitioned $x$, find the length distribution $l$ of an optimal variable length code:

$$l(x') = -\log p(x') \quad (6)$$

3) given the representatives $x'$, find the best partitioning by assigning each data point x to the representative $x'$ that minimizes:

$$d_{KL}^{(y)}(x,x') + \lambda \cdot l(x') \quad (7)$$

where $\lambda$ is a Lagrange parameter balancing the costs of distortion and transmission.

By picking $\lambda$, we can choose between a course quantization leading to a short description $x'$ of the features, that has a relatively high loss of information concerning $y$ ($\lambda$ large), or a longer description resulting from a finer quantization that contains much more detail concerning $y$ ($\lambda$ small). Of course, the information-theoretic interesting aspect is to find the largest possible $I(X';Y)$ for a given average description length $H(X')$. Note that the extended Lloyd algorithm can be applied both on a scalar $x$ (scalar quantization), and on a vector $x$ (vector quantization).

Typically, the modified Lloyd algorithm starts off in a space $x$ of high-resolution quantized features, where in each bin of the space the relevant probabilities $p(x|y)$, $p(x)$ and $p(y|x)$ can be defined, estimated or computed. The algorithm subsequently performs an optimal partitioning of this space into $K$ regions $x'$ by a clustering of the high-resolution bins $x$. It seems that both the number $K$ and a good initial location of $x'$ must be found experimentally.

# A Card-Based Cryptographic Protocol and a Logic Puzzle

Mitsugu Iwamoto
The University of Electro-Communications, Japan
Email: mitsugu@uec.ac.jp

ABSTRACT

*Multi-Party Computation* (MPC) enables mutually distrusted $n$ parties to compute a function on their $n$ inputs securely. It is well known that MPC is important for applications of cryptography, e.g., for computing statistics of cloud data. MPC is also fascinating for understanding and enjoying cryptography since it can be implemented by using a deck of playing cards [1], which is called "*card-based cryptographic protocol*". We address in this talk a card-based cryptographic protocol for secure comparison of two numbers, called *millionaires' problem*, which is a traditional problem in MPC initiated by Yao's seminal work [2].

Our original motivation presented in [3] is to propose efficient card-based cryptographic protocols for millionaires' problem in terms of MPC, i.e., communication complexities and number of cards used in the protocol. In this talk, however, we rather focus on an interesting relation between the proposed protocol and a famous logic puzzle "*The fork in the road*" (e.g., see [4]), which was firstly pointed in [3].

In the proposed protocol, we assume that the wealth of Alice and Bob is represented by bits, and they execute bitwise comparisons for their wealth. In each bitwise comparison, Alice sends her bit or its negation to Bob without telling him which one of them is sent. Even if she tells a lie on her bit, it is possible to compare her bit with Bob's one by introducing a technique borrowed from "The fork in the road."

ACKNOWLEDGEMENT

This work was supported by JSPS KAKENHI Grant Numbers JP26420345, JP15H02710, and JP17H01752.

# Author Index

# Proceedings of Workshop AEW10 Concepts in Information Theory and Communications